# Extensive simulation studies on the reconstructed image resolution of a position sensitive detector based on pixelated CdTe crystals

K. Zachariadou, K. Karafasoulis, I. Kaissas, S. Seferlis, C. Lambropoulos, D. Loukas, C. Potiriadis

*Abstract*–We present results on the reconstructed image resolution of a position sensitive radiation instrument (COCAE) based on extensive simulation studies. The reconstructed image resolution has been investigated in a wide range of incident photon energies emitted by point-like sources located at different source-to-detector distances on and off the detector's symmetry axis. The ability of the detector to distinguish multiple radioactive sources observed simultaneously is investigating by simulating point-like sources of different energies located on and off the detector's symmetry axis and at different positions.

I. INTRODUCTION

THE COCAE instrument [1] is a portable detecting system under development aimed to be used to accurately detect the position as well as the energy of radioactive sources in a broad energy range up to ~2 MeV. Among its applications could be the security inspections at the borders and the detection of radioactive sources into scrap metals at recycling factories.

For the source localization task, COCAE exploits the Compton scattering imaging [2], a technique widely used in many fields such as nuclear medicine, astrophysics and recently counterterrorism. Instruments like COCAE that exploit the Compton imaging technique deduce the energy of the incident gamma ray photons as well as their origin within a cone, by measuring the energy depositions and the positions of the Compton scattering interactions recorded in the detector.

Manuscript received November 15, 2011. This work was supported by the collaborative Project COCAE SEC-218000 of the European Community's Seventh Framework Program.

K. Zachariadou is with the Greek Atomic Energy Commission, Agia Paraskevi, Attiki, Greece and with the Technological Educational Institute of Piraeus, P. Ralli–Thivon-Athens, Greece (e-mail: zacharia@inp.demokritos.gr).

K. Karafasoulis is with the Greek Atomic Energy Commission, Agia Paraskevi, Attiki, Greece and with the Hellenic Army Academy, Vari, Attiki Greece (e-mail: ckaraf@gmail.com).

I. Kaissas is with the Greek Atomic Energy Commission, Agia Paraskevi, Attiki, Greece (e-mail: ikaissas@eeae.gr).

S. Seferlis is with the Greek Atomic Energy Commission, Agia Paraskevi, Attiki, Greece (e-mail: stsefer@eeae.gr).

C.P. Lambropoulos is with the Technological Educational Institute of Chalkida, Psahna – Evia, 34400 Greece (e-mail: lambrop@teihal.gr).

D. Loukas is with the Institute of Nuclear Physics, National Center for Scientific Research, Agia Paraskevi, Attiki, Greece (e-mail: loukas@inp.demokritos.gr).

C. Potiriadis is with the Greek Atomic Energy Commission, Agia Paraskevi, Attiki, Greece (e-mail: cpot@eeae.gr).

Successive interactions of an incident photon create an overlapping cone. The intersection of the cones corresponding to different incident photons determines the source location. In principle, three cones are sufficient to reconstruct the image of a point source. In practice, due to measurement errors and incomplete photon absorption, a large number of reconstructed cones are needed to derive the source location accurately.

The instrument consists of ten parallel planar layers made of pixelated Cadmium Telluride (CdTe) crystals occupying an area of 4cmx4cm, placed 2cm apart from each other. Each detecting layer consists of a two-dimensional array of pixels (100x100) of 400μm pitch, bump-bonded on a two-dimensional array of silicon readout CMOS circuits. Both pixels and readout arrays are on top of an $Al_2O_3$ supporting printed circuit board layer.

For the Monte Carlo simulation studies completed for the development of the COCAE instrument an open-source object-oriented software library (MEGAlib [3]) has been used, which provides an interface to the Geant4 [4] toolkit that simulates the passage of particles through matter.

Important performance parameters of the COCAE instrument such as its detecting efficiency and angular resolution have been studied by Monte Carlo simulation [5]. Since the determination of the correct sequence of photon interactions with the detector via Compton scattering strongly affects the detector's efficiency of evaluating the direction of the incident photons, various techniques [6] (that exploit the kinematical and geometrical information of Compton scattering events as well as statistical criteria) have been extensively studied in order to select the best one [7]. Furthermore, the overall efficiency of event reconstruction has been evaluated in a wide range of initial photon energies [8].

The main task of the present work is the estimation of the instrument's reconstructed image resolution for point-like sources as well as the investigation of the instrument's ability to distinguish multiple point-like sources.

II. RECONSTRUCTED IMAGE RESOLUTION

The images of point-like radioactive sources have been reconstructed by applying the List Mode Maximum Likelihood Expectation Maximization (LM-MLEM) imaging algorithm [9]. According to this algorithm the image of a point source is generated by projecting each Compton event cone into an imaging projection sphere and then by performing successive iterations on the back-projected image in order to find the reconstructed image distribution with the highest likelihood of having produced the observed data.

Fig. 1 depicts the reconstructed image for the case of an 800keV point-like source located at 50cm distance from the detector's center at azimuth angle $\theta = 26.56^0$ and inclination angle $\phi = 0^0$ (where $\theta = 0^0$ corresponds to the detector symmetry axis). The azimuth and inclination profiles of the reconstructed image are shown in Fig. 2.

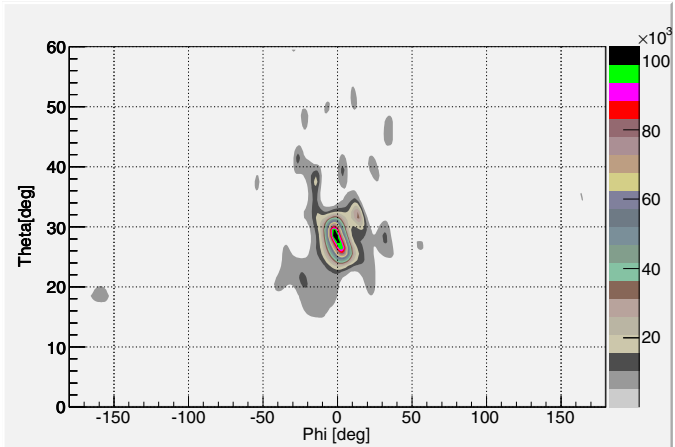

Fig. 1 Reconstructed image of an 800 keV point-like source located at θ=26.56°, φ=0°.

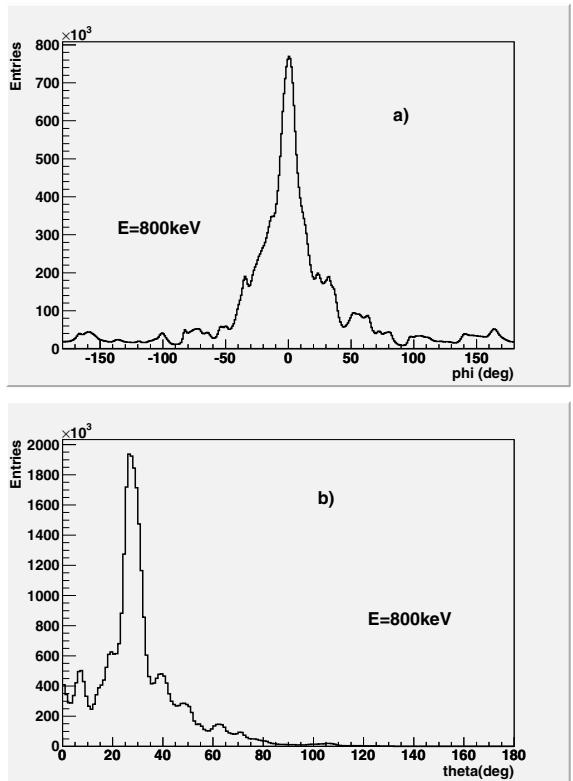

Fig. 2 a) azimuth and b) inclination profiles of the reconstructed image of an 800keV point-like source located at θ=26.56°, φ=0°.

The reconstructed image resolution of the COCAE instrument is defined as the FWHM of the reconstructed image distribution measured in steradian (sr) and it has been studied by considering two case conditions: a) point-like sources located on the detector's symmetry axis (z) and b) point-like sources located off the detector's symmetry axis.

For the case of on-axis sources, results demonstrate that the imaging resolution varies from ~2.5x10$^{-3}$ sr (for source-to-detector distances ~50cm) down to ~0.5x10$^{-3}$ sr for point-like sources located at distances greater than ~1m (Fig. 3). For the reconstructed image resolution evaluation under various conditions (photon's energy, source-to-detector distance and orientation), the same number of photons-detecting materials interactions has been assumed. Experimentally this assumption can be achieved by increasing the acquisition time as a function of the source-to-detector distance.

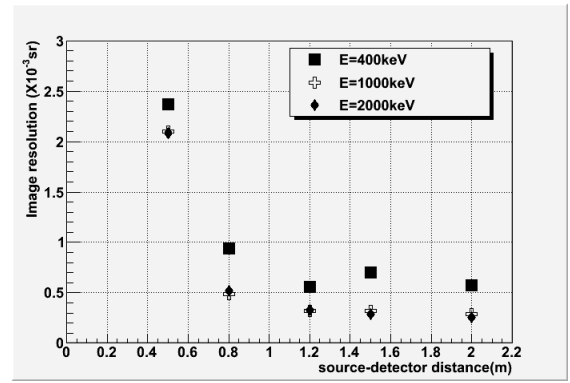

Fig. 3 Reconstructed image resolution for point-like sources located on the detector's symmetry axis emitting 400 keV, 1000 keV and 2000 keV photons, as a function of source-to-detector distance.

Moreover, it can be seen from Fig. 4 that for an arbitrary source-to-detector distance (z=80cm) and for point-like sources emitting photons with energies from 400keV to 2000keV the reconstructed image resolution is ~0.5x10$^{-3}$sr.

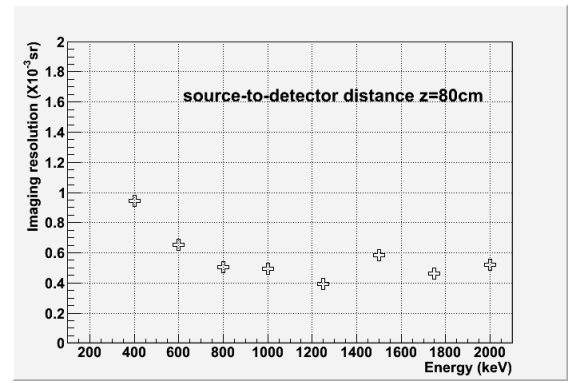

Fig. 4. Reconstructed image resolution for point like sources located on the detector's symmetry axis as a function the incident gamma ray energy, for an arbitrary source-to-detector distance (z=80cm).

For the case of point-like sources located off the detector's symmetry axis, it can be seen (Fig.5) that the reconstructed image resolution is less than ~4x10$^{-3}$sr, for point-like source emitting photons with energies from 600keV to 2000keV.

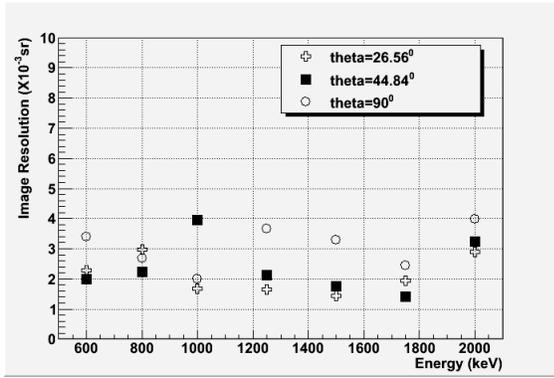

Fig. 5 Reconstructed image resolution as a function of the incident photon energy, for different inclination angles.

Fig. 6 shows the reconstructed imaging resolution versus the inclination angle for 600keV, 1000keV and 2000keV sources.

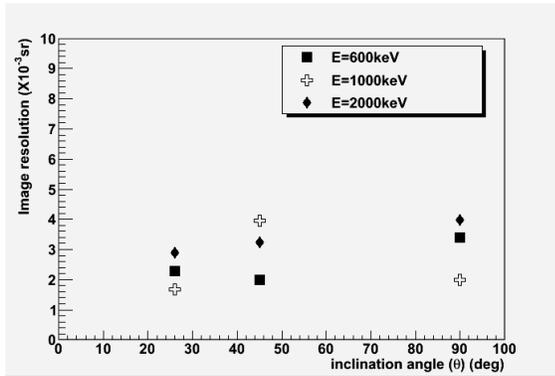

Fig. 6. Reconstructed image resolution, as a function of the inclination angle, for different incident photon energies.

### III. MULTIPLE SOURCES DISCRIMINATION

The ability of the COCAE detector to resolve two point-like sources spatially separated has been studied. Two source conditions have been considered a) point-like sources located at different source-to-detector distances and b) point-like sources located at the same source-to-detector distances. Both sources emit photons with the same energy and are located on and off the detector's symmetry axis (Fig. 7).

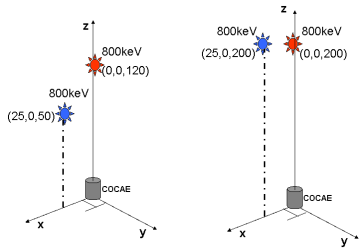

Fig. 7: Geometry for studying the two-source discrimination ability of the COCAE instrument, for point-like sources located at different distances from the detector (left) and for point-like sources located at the same distance from the detector (right).

For the image reconstruction the List Mode Maximum Likelihood Expectation Maximization (LM-MLEM) imaging algorithm has applied on the same number of incident photons.

Shown in Fig.8 is the case of two 800keV point-like sources located at an arbitrary source-to detector distance on and off the detector's symmetry axis (x=0, y=0, z=120 cm) and (x=25, y=0, z=50 cm). The case of sources located at the same (arbitrary) source-to-detector distance (z=200cm) is depicted in Fig.9.

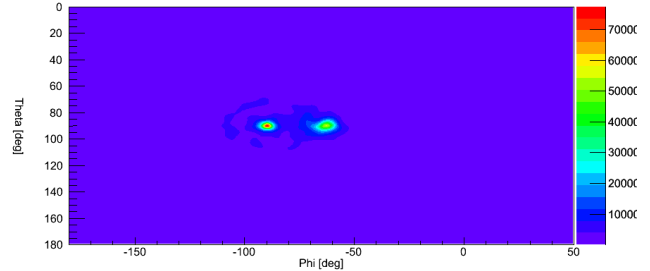

Fig. 8: Reconstructed images for 800keV sources located at (x=25, y=0, z=50 cm) and (x=0, y=0, z=120 cm)

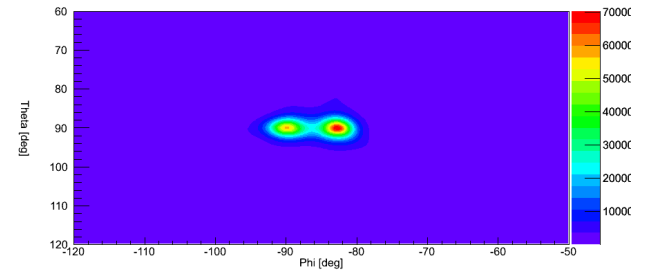

Fig. 9. Reconstructed images for 800keV sources located at the same source-to-detector distance: (x=25, y=0, z=200 cm) and (x=0, y=0, z=200 cm)

Furthermore, Fig. 10 depicts the case of two sources with angular separation less than the reconstructed image resolution of the COCAE instrument.

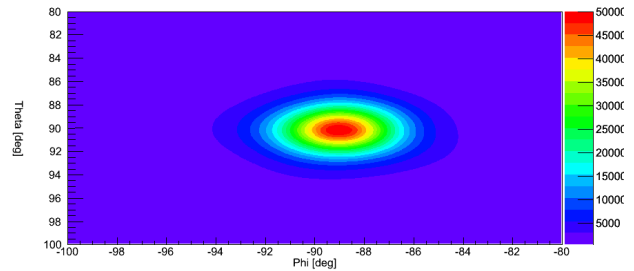

Fig. 10. The two 800keV radioactive sources located at (x=7, y=0,z=200 cm) and (x=0, y=0, z=200 cm) are out of the COCAE detector's discrimination limit.

### IV. CONCLUSION

The imaging resolution of the portable detector (COCAE) based on pixelated CdTe crystals has been studied by applying the List Mode Maximum Likelihood Expectation Maximization (LM-MLEM) imaging algorithm on a large number of simulated events.

The reconstructed image resolution has been studied for two cases: on and off the detector's symmetry axis. Results demonstrate that the resolution is worse for off-axis sources, short source-to-detector distances as well as for sources emitting low energy photons. The COCAE's detector estimated reconstructed image resolution is less than ~$4 \times 10^{-3}$ sr.

Furthermore, the ability of the COCAE's detector to resolve two point-like sources spatially separated has been studied for same energy point-like sources located at different source-to-detector distances as well as for point-like source located at the same source-to-detector distances.

Further simulation studies are underway to explore the reconstructed image resolution for evaluating the reconstructed image resolution in the case of point-like sources emitting photons with different energies as well as for estimating the detector's minimum detection limit under all the above mentioned various conditions.